%% file: main.tex
\documentclass[aps,pre,onecolumn,superscriptaddress,nofootinbib]{revtex4-2}

\usepackage[utf8]{inputenc}
\usepackage[T1]{fontenc}
\usepackage{amsmath,amssymb,amsfonts}
\usepackage{graphicx}

\usepackage{physics}
\usepackage{float}
\usepackage{lipsum}
\usepackage{comment}
\usepackage{subcaption}
\usepackage[margin=1in]{geometry}
\usepackage[hidelinks]{hyperref}

\begin{document}

\input{chapters/Foreword}% عنوان
\input{chapters/abstract}% چکیده
\maketitle

\input{chapters/introduction}%مقدمه 
\input{chapters/methods}%روش کار 
\input{chapters/results}%نتایج

\input{chapters/conclusions}% بحث و پیشنهادات

	% Final insights, future work...
	
	\begin{acknowledgments}
	This work benefited from editorial and technical assistance provided by ChatGPT, developed by OpenAI.
	\end{acknowledgments}
	
	\bibliographystyle{apsrev4-2}
	\bibliography{references}
	
\end{document}

%% file: chapters/Foreword.tex
\title{A Neuronal Model at the Edge of Criticality: An Ising-Inspired Approach to Brain Dynamics}

\author{Sajedeh Sarmastani}
\affiliation{Department of Physics, Tarbiat Modares University, Tehran, Iran}

\author{Maliheh Ghodrat}
\affiliation{Department of Physics, Tarbiat Modares University, Tehran, Iran}

\author{Yousef Jamali}
\thanks{Corresponding author}
\email{y.jamali@modares.ac.ir}
\affiliation{Department of Mathematics, Tarbiat Modares University, Tehran, Iran}

\date{\today}

%% file: chapters/abstract.tex
\begin{abstract}
	We present a neuronal network model inspired by the Ising model, where each neuron is a binary spin ($s_i = \pm1$) interacting with its  neighbors on a 2D lattice. Updates are asynchronous and follow Metropolis dynamics, with a temperature-like parameter $T$ introducing stochasticity.
	
	To incorporate physiological realism, each neuron includes fixed on/off durations, mimicking the refractory period found in real neurons. These counters prevent immediate reactivation, adding biologically grounded timing constraints to the model.
	
	As $T$ varies, the network transitions from  asynchronous to synchronised activity. Near a critical point $T_c$, we observe hallmarks of criticality: heightened fluctuations, long-range correlations, and increased sensitivity. These features resemble patterns found in cortical recordings, supporting the hypothesis that the brain operates near criticality for optimal information processing.
	
	This simplified model demonstrates how basic spin interactions and physiological constraints can yield complex, emergent behavior, offering a useful tool for studying criticality in neural systems through statistical physics.
	
\end{abstract}

%% file: chapters/introduction.tex
\section{Introduction}

Understanding the complex dynamics of neural systems remains a central challenge in neuroscience. One compelling hypothesis suggests that the brain operates near a critical point \_ a phase transition between order and disorder —that enables optimal computational performance. At this critical regime, the brain is thought to maximize dynamic range, facilitate efficient information transmission, and enhance sensitivity to external stimuli \cite{RN33,RN72,RN18,RN35,RN73,RN54,RN52,RN35,RN48,RN44,RN79,RN18,RN80}.

The Ising model, initially developed to describe ferromagnetic phase transitions, has emerged as a foundational framework for exploring complex systems with local interactions and emergent global order. By mapping binary spin variables to neuronal activity states, the Ising model offers a simplified yet powerful representation of collective dynamics in neural networks. In this representation, each neuron is modeled as a binary unit (active/inactive), and large-scale network behavior arises from local coupling rules \cite{RN77,RN78,RN4,RN47}.

In the classical Ising model,temperature $T$  quantifies thermal noise that competes with interaction energy \cite{RN28,RN7}. In our model, $T$  analogously represents physiological noise sources influencing spike generation. These include the stochastic gating of ion channels \cite{RN81}, probabilistic synaptic vesicle release \cite{RN82,RN83}, and spontaneous background activity \cite{RN84}. Additionally, neuromodulators like dopamine, serotonin, and acetylcholine modulate neuronal excitability \cite{RN86,RN85}, effectively altering the system's 'temperature'. Higher values of $T$  correspond to heightened stochastic influences that can drive neural activation even in the absence of strong synaptic input.

However, conventional Ising-based neural models typically overlook key biological features, such as the \textit{refractory period} \_a brief, intrinsic time window following a neuron's activation during which it cannot immediately fire again \cite{RN33}. The refractory period plays a vital role in shaping temporal patterns, rhythmic synchronization, and the stability of neural codes. Including such physiological constraints is crucial for creating more accurate models of cortical dynamics \cite{RN69}.

In this study, we propose a biologically grounded extension to the classical Ising model by incorporating fixed ON and OFF durations for each neuron, thereby mimicking the refractory behavior observed in cortical neurons. Neurons are placed on a two-dimensional lattice with enhanced local connectivity (28 neighbors), and their state updates are governed by asynchronous Metropolis dynamics. The imposed ON/OFF cycles prevent immediate reactivation, adding a critical temporal constraint that better captures real neuronal behavior.

Within this extended framework, we investigate the emergence of critical phenomena in the modified network.By increasing $T$ , the system transitions from disordered, asynchronous state to ordered , synchronized  behavior. At a critical point $T_c$, the balance between noise and connectivity generates a regime that the network exhibits features characteristic of second-order phase transitions: peak fluctuations in neuronal activity, maximal responsiveness and complexity , long-range spatial and temporal correlations — (reminiscent of the hypothesised critical point of the cerebral cortex).

Our results demonstrate that even minimal modifications to traditional spin-based models, such as incorporating a refractory period, enable the emergence of rich, biologically relevant behaviours. This work strengthens the evidence supporting the critical brain hypothesis and provides a tractable, interpretable framework for studying criticality in neural systems.

%% file: chapters/methods.tex
\section{Model}

Our neuronal model is a biologically-inspired modification of the classical Ising model, where each neuron is represented by a spin-like variable $s_i \in \{-1,+1\}$ denoting inactive and active states, respectively. The network comprises $l \times l$ = N neurons arranged on a two-dimensional lattice with periodic boundary conditions and enhanced local connectivity, including 28 neighbors (Neighborhood with a radius of 3 neurons).

The system's energy is described by the Hamiltonian:
\begin{equation}
	\label{eq:hamiltonian}
	H = -\sum_{\langle ij \rangle} J_{ij} s_i s_j
\end{equation}
where $J_{ij}$ represents the synaptic coupling between neurons $i$ and $j$. We assume uniform coupling strength $J_{ij} = J = 1$ for all connected neuron pairs.

To incorporate physiological neuron behavior, we integrate a time persistence mechanism into the model. 
As observed in the behavior of real neurons, when an external stimulus reaches a specific threshold, the probability of neuron activation increases. 
When a neuron activates, it remains active for a defined number of ``on time'' steps. After this period, it transitions to an inactive state for a fixed number of ``off time'' steps, analogous to the biological refractory period. Neurons become eligible for state transitions again only after completing both phases. 

\begin{figure}[htb]
	\centering
	\includegraphics[width=0.7\linewidth]{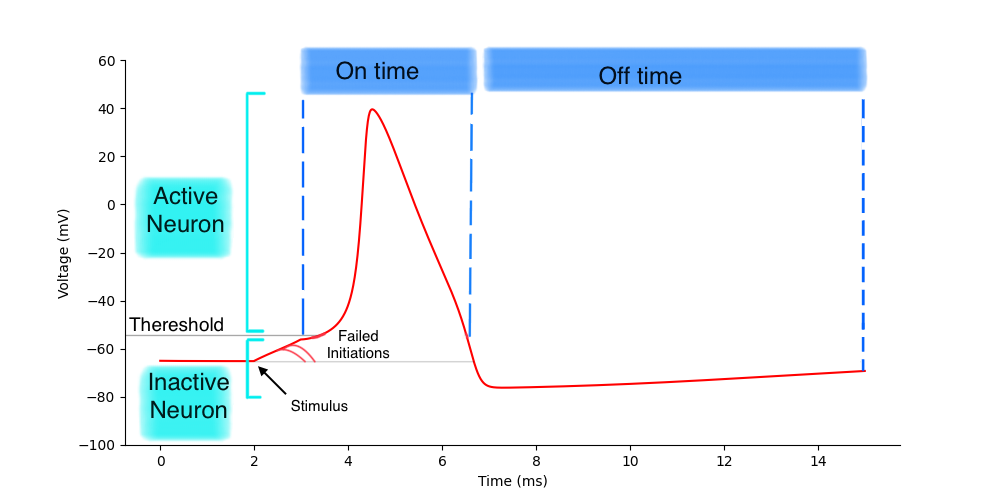}
	\caption{In a real neuron, this corresponds to the action potential. In the simulated model based on a real neuron, once an inactive neuron becomes active, it remains active for a fixed period ($On\ time$) and then inactive for another period ($Off\ time$), after which the probability of reactivation is evaluated using the Metropolis algorithm.  }
	\label{fig:action-potential2}
\end{figure}

This updating follows the stochastic rule provided by the Metropolis algorithm, mimicking threshold activation and neuronal recovery dynamics.

\begin{equation}
	\label{eq:metropolis_neuron}
	P(s_i = -1 \rightarrow s_i = 1) = 
	\begin{cases}
		1, & \text{if } \Delta H_i \leq 0, \\[8pt]
		\exp\left(-\frac{\Delta H_i}{T}\right), & \text{if } \Delta H_i > 0,
	\end{cases}
\end{equation}

 Activation probability depends on the energy change $\Delta H$ (Role of active neighbors to activation of neuron)  and noise  or external stimuli  $T$ (Role of neuromodulators in the activation of neuron)
 , following the exponential rule $e^{-\Delta H / T}$. 
 
 If $\Delta H_i$ is negative (most neighbors are active), activation is  certain.  Conversely, if $\Delta H$ is positive  but noise $T$ is substantial (There is a large amount of excitatory neuromodulators), probabilistic activation is still possible with $e^{-\Delta H_i / T}$, Otherwise, the neuron remains inactive.

 In our implementation, all neuron spins are updated sequentially at each time step and  neurons are excitatory and solely activate neighboring neurons. The Metropolis algorithm exclusively determines the transition from inactive to active states.

Simulations were conducted across various temperatures to explore phase transitions in neuronal network activity. 
Average neuronal activity 
\begin{equation}
	\label{eq:magnetization}
	M = \frac{1}{N} \sum_{i=1}^{N} s_i
\end{equation}, serves as the order parameter , analogous to the role of magnetization in the Ising model.

This formulation captures crucial aspects of neuronal dynamics, including synchronization, refractory behavior, and emergent criticality within complex neural systems.

%% file: chapters/results.tex
\section{Results}
The proposed neuronal network model, inspired by the physiology of neurons and the Ising model , was implemented using the Metropolis algorithm to probabilistically determine neuron activation . The model exhibits different behaviors depending on the temperature parameter ($T$), particularly in subcritical, critical, and supercritical regimes .
\subsection{Network Activity}
The network activity, denoted by $M$, is defined as the average state of neurons  $s_i$ , at each time step by 
\ref{eq:magnetization}
where $s_i$ is the state of neuron $i$. This parameter 'M' , ranges from -1 (all neurons off) to 1 (all neurons on), with values close to zero indicating random activity .

Analyzing the network activity over time reveals different behaviors across temperature regimes including subcritical, critical, and supercritical.

\begin{figure}[H]
	\centering
	\includegraphics[width=\linewidth]{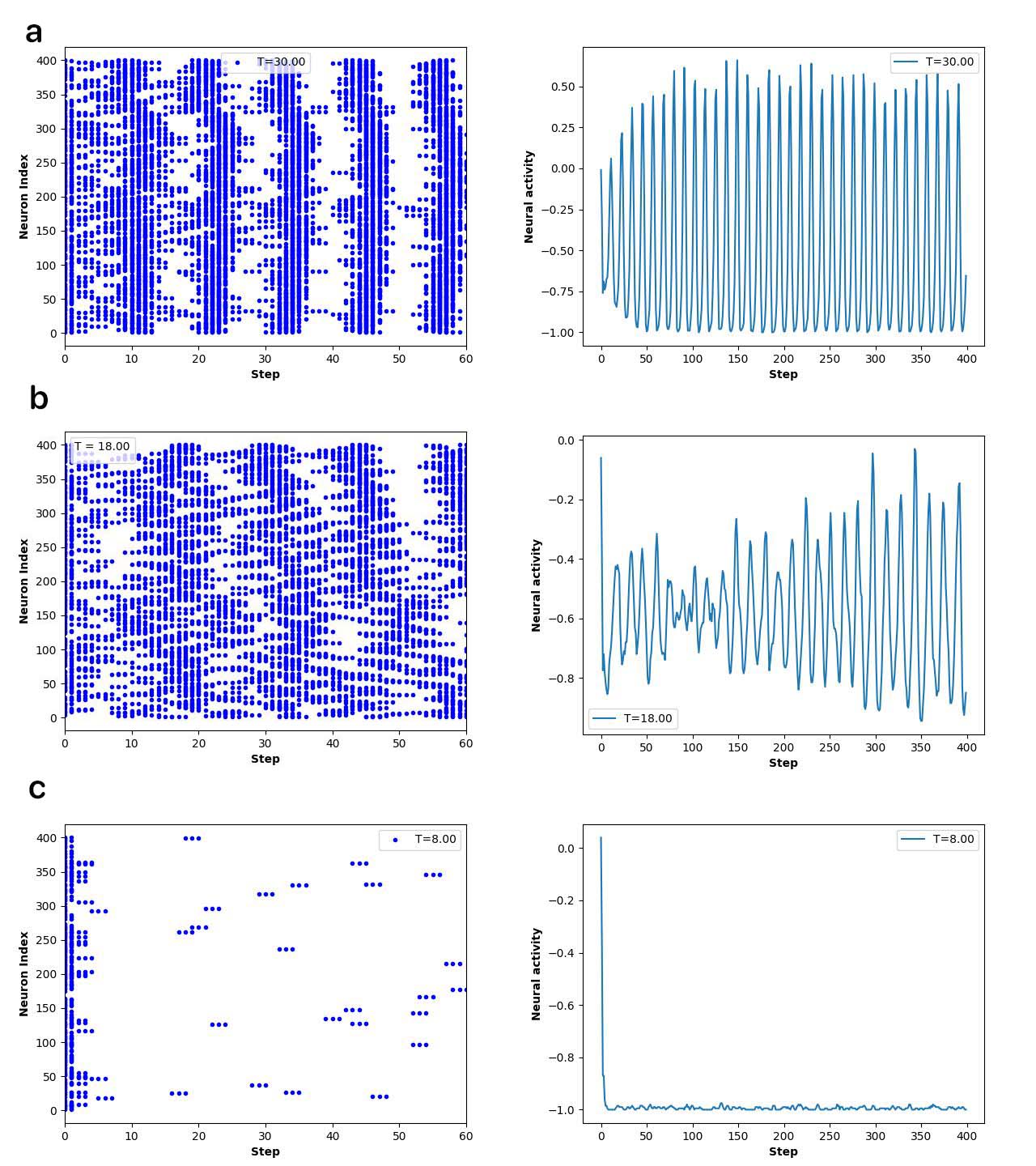}
	\caption{Neuronal dynamics across different temperature regimes. Each row corresponds to a specific temperature: (a) supercritical ($T = 30$), (b) critical ($T = 18$), and (c) subcritical ($T = 8$). For each temperature, the left panel shows the raster plot of neural activity (black dots represent active neurons), and the right panel shows the corresponding average network activity over time. In the supercritical regime, the system exhibits synchronized oscillatory activity. In the subcritical regime, the system gradually settles into an inactive state (all spins $= -1$). At the critical temperature, a combination of ordered and disordered behavior emerges, reflecting enhanced dynamical complexity and long-range spatiotemporal correlations}
		\label{fig:raster_mag_plot}
\end{figure}

\begin{itemize}
	\item At \textbf{supercritical }temperatures, such as ($T=30$) the network shows coordinated behavior . Neurons turn on and off together, exhibiting oscillations with a specific period determined by the defined on and off times for neurons . Figure a-\ref{fig:raster_mag_plot}  right and left the network activity over time and raster plot at $T=30$ .
	
	\item At the \textbf{critical }temperature such as ($T=18$), the network exhibits a more complex behavior, representing a combination of coordinated and uncoordinated activity . This is visible in the raster plot in left b-\ref{fig:raster_mag_plot} , and the  network activity over time in  Figure right b-\ref{fig:raster_mag_plot} .

	\item At \textbf{subcritical }temperatures, such as ($T=8$), the network gradually settles into an inactive state. Neurons activate sparsely and mostly remain off, resulting in low overall activity. This is evident from both the network activity and the raster plot in Figure c-\ref{fig:raster_mag_plot}, where the system displays disordered, asynchronous firing and lacks global coordination.
\end{itemize}

\subsection{Phase transition}
\begin{figure}[htb]
	\centering
	\includegraphics[width=0.6\textwidth]{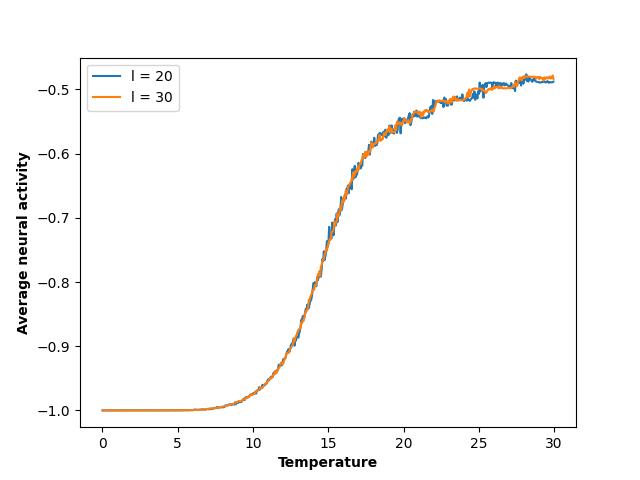}
	\caption{
	Mean neuronal activity $\bar{M}$ as a function of temperature $T$ for various network sizes $N$. A smooth transition from disordered (asynchronous) to ordered (synchronous) states is observed, indicating a continuous (second-order) phase transition characteristic of critical phenomena.
	}
	\label{fig:Average_Magnetization_All_N}
\end{figure}

The mean network activity, $\bar{M}$, is calculated by averaging the activity $M_i$ over time steps, excluding data before equilibrium is reached.
An investigation of the system's order parameter (mean neuronal activity) reveals a continuous phase transition from disorder to order with increasing temperature . This continuous change in the mean network activity with increasing temperature (control parameter) is a sign of a second-type phase transition in the system . This is visualized in Figure \ref{fig:Average_Magnetization_All_N}, which shows the mean activity at different temperatures for different network sizes .

%%%%%%%%%%%%%%%%%%%%%%%%
\subsection{System Energy}

The system's energy ($E$) is calculated using Equation
$E =  \frac{-J }{N} \sum_{\langle i,j \rangle} S_i S_j$
where the sum is over neighboring pairs.

At low temperatures, the probability of neurons turning on decreases, and the system tends towards an inactive state. Consequently, the total system energy approaches its minimum value (-1) . As temperature increases, the probability of neurons turning on via the Metropolis algorithm increases, causing the network to move away from the completely inactive state, and the system's energy increases . In an intermediate temperature range, the system experiences a combination of irregular and regular on and off state, and the energy approaches an average value . At high temperatures, while random behavior is expected due to strong noise, two factors prevent the energy from reaching a completely high value: strong interactions with neighbors and the limited on/off time duration of neurons . As a result, the system maintains some degree of correlation even at high temperatures, and the energy approaches a nearly constant value dependent on the neuron's on and off times. 
%%%%%%%%%%%%%%%%%%%%%%%%
\subsection{Dynamic Capacity and Neuron Sensitivity}

Building on the concept of heat capacity in the classical Ising model, we define the dynamic capacity ($C$) of the neuronal network to quantify its energy fluctuation response:
\begin{equation}
	C = \frac{1}{k_B T^2 } (\langle E^2 \rangle - \langle E \rangle^2),
\end{equation}
where $T$ is the temperature, $k_B$ is Boltzmann's constant (set to $1$), $\langle E \rangle$ is the average energy, and $\langle E^2 \rangle$ is the mean squared energy. This expression captures the variance of energy, representing the system's sensitivity to thermal noise.

\begin{figure}[htb]
	\centering
	\includegraphics[width=\textwidth]{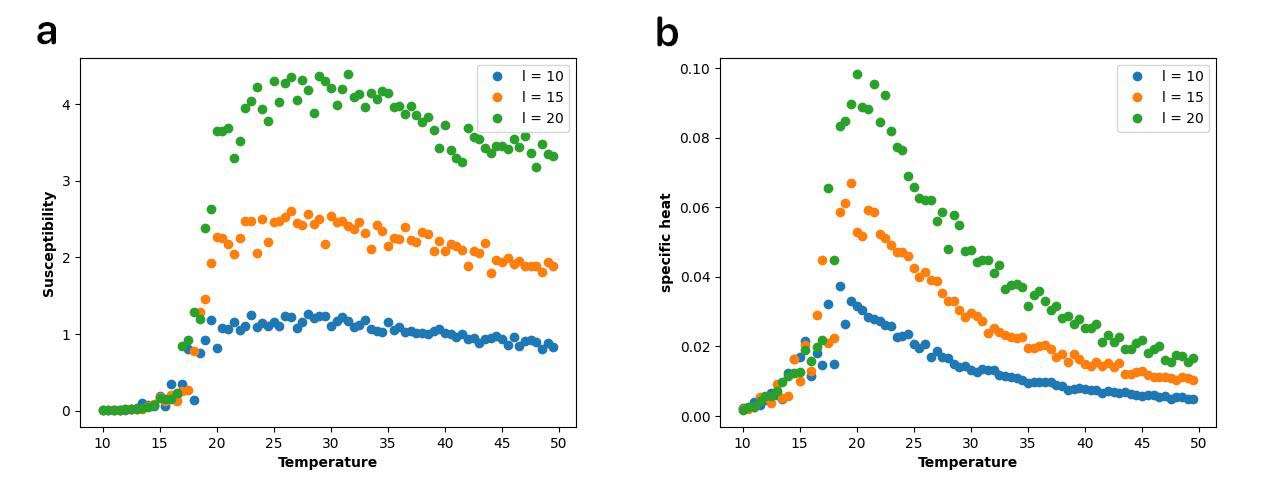}
	\caption{(a) Temperature dependence of neuronal susceptibility ($\chi$) for network sizes $L = 10$, $15$, and $20$. Susceptibility peaks more sharply and at higher values as system size increases, indicating enhanced responsiveness near the critical point.
		(b) Temperature dependence of specific heat ($C$) for the same network sizes. The specific heat also peaks around the critical temperature , with larger systems exhibiting sharper and higher peaks, consistent with second-order phase transition behavior.}
	\label{fig:xc}
\end{figure}

In parallel, we define neuron sensitivity ($\chi$), analogous to magnetic susceptibility, as:
\begin{equation}
	\chi = \frac{1}{k_B T } (\langle M^2 \rangle - \langle M \rangle^2),
\end{equation}
where $\langle M \rangle$ is the mean network activity (mean magnetization), and $\langle M^2 \rangle$ is the mean squared magnetization. This metric quantifies the variance in neuronal activity, measuring the network's responsiveness to small external perturbations.
Figurea in a-\ref{fig:xc} illustrates the temperature dependence of neuronal sensitivity ($\chi$), while Figureb in b-\ref{fig:xc} presents the corresponding behavior of dynamic capacity ($C$).

At low temperatures, both $C$ and $\chi$ remain minimal, indicating a stable regime with low fluctuations in neuronal activity and energy. As the temperature approaches the critical threshold $T_c$, both quantities exhibit a sharp increase, peaking precisely at $T_c$. This pronounced peak becomes sharper and higher with increasing network size, a hallmark of a second-order (continuous) phase transition.

Beyond the critical point, both $C$ and $\chi$ decline, reflecting the system's return to a ordered phase dominated by thermal noise and reduced responsiveness.

The peak in $\chi$ near $T_c$ highlights enhanced sensitivity to external stimuli, mirroring the behavior of spin systems near magnetic criticality. Concurrently, the peak in $C$ signifies maximal energy fluctuations, capturing the network’s elevated response to stochastic perturbations.

Together, these critical peaks affirm that the system undergoes a continuous phase transition analogous to that of the classical Ising model. This behavior supports the hypothesis that biological neural systems may operate near criticality to optimize their computational efficiency, dynamic range, and adaptability.

%%%%%%%%%%%%%%%%%%%%%%%%%
\subsection{Correlation}
\subsubsection{Time Correlation}

The time correlation or autocorrelation function (ACF) of the neural network activity is calculated using the following expression \cite{RN9}, analogous to time correlation analysis in the Ising model:

\begin{equation}
	\mathrm{ACF}(\tau) = \frac{ \langle  \left( M - \langle M \rangle \right) \left( M(\tau) - \langle M \rangle \right)\rangle}{  \langle M^2 \rangle - \langle M \rangle^2 }
\end{equation}
where $M(t)$ is the network activity  at time $t$, $\tau$ is the time delay and $\langle M \rangle$ is the average neural activity.

 \begin{figure}[htb]
\centering
\includegraphics[width=\linewidth ]{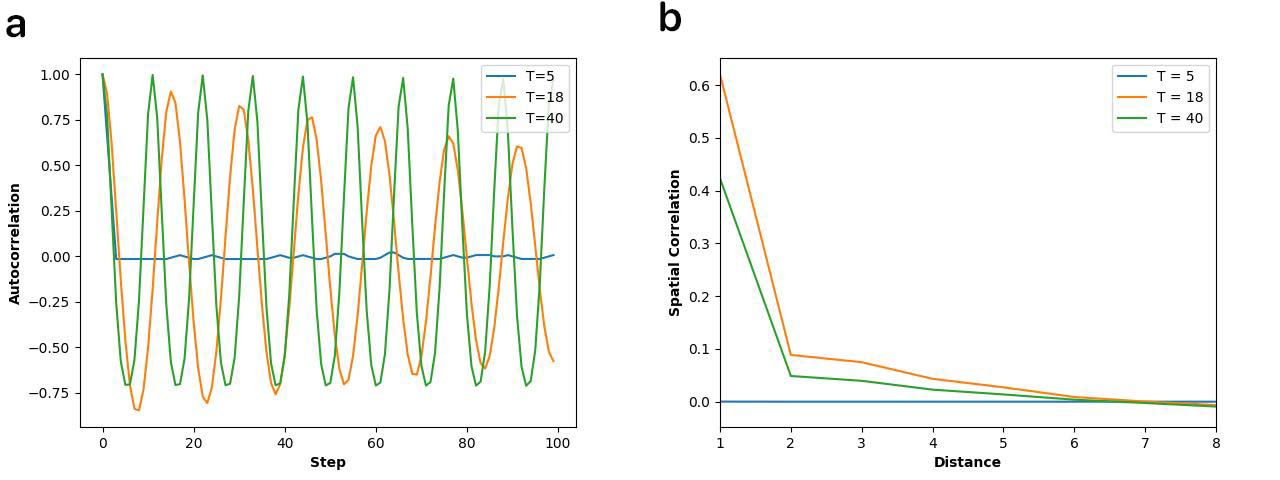}
\caption{}
\label{fig:ac}
\end{figure}

As illustrated in Figure \ref{fig:ac}-a, the ACF exhibits distinct behaviors depending on the temperature:
\begin{itemize}
	\item \textbf{Subcritical regime} ($T=5$): The autocorrelation fluctuates around zero, indicating weak memory and information transmission of the system.
	\item \textbf{Critical regime} ($T = T_c = 18$): The ACF oscillates with reduced amplitude and decays slowly, indicating the emergence of long-range temporal correlations. The slow decay implies long memory and a prolonged relaxation time to equilibrium.
	
	\item \textbf{Supercritical regime} ($T=40$): The ACF shows strong oscillations with fixed frequency due to high thermal noise. In fact, the oscillations of neural activities are mirrored in the autocorrelation, resulting in an oscillatory profile.
\end{itemize}

At zero delay ($\tau = 0$), $\mathrm{ACF}(0) = 1$ by definition, representing full correlation.

These findings underscore how temperature affects memory and correlation in the neural network model, highlighting criticality as the point where temporal structure is richest and most persistent.

%%%%%%%%%%%%%%%%%

\subsubsection{ Spatial Correlation}

Spatial correlation quantifies how the state of one neuron is statistically dependent on the state of another neuron located at a distance $r$ away. This is measured using the normalized two-point correlation function:
\begin{equation}
	C(r) = \frac{\langle S S_{r} \rangle - \langle S \rangle \langle S_{r} \rangle}{\langle S^2 \rangle - \langle S\rangle^2}
\end{equation}
where $S$ is the state of neuron , and $S_{r}$ denotes the state of the neuron located a distance $r$ . The angle brackets denote averages taken over time and over all neuron pairs separated by distance $r$.

As illustrated in Figure \ref{fig:ac}-b, the spatial correlation $C(r)$ generally decreases as the distance $r$ increases. This decay characterizes how quickly the influence of a neuron's state vanishes with spatial separation.

Notably, near the critical temperature ($T=18$), the decay of $C(r)$ is markedly slower, and correlations persist over larger distances. This indicates that neurons influence not only their immediate neighbours but also more distant ones, reflecting the emergence of long-range spatial correlation typical at criticality.

These results confirm that the neuronal network exhibits extended spatial dependencies near $T_c$, consistent with the behavior expected in systems undergoing second-order phase transitions.
%%%%%%%%%%%%%%%%%%%%%%%%%

\subsection{Cluster Size Distribution}

\begin{figure}[htb]
	\centering
	\begin{subfigure}[b]{0.49\textwidth}
		\includegraphics[width=\textwidth]{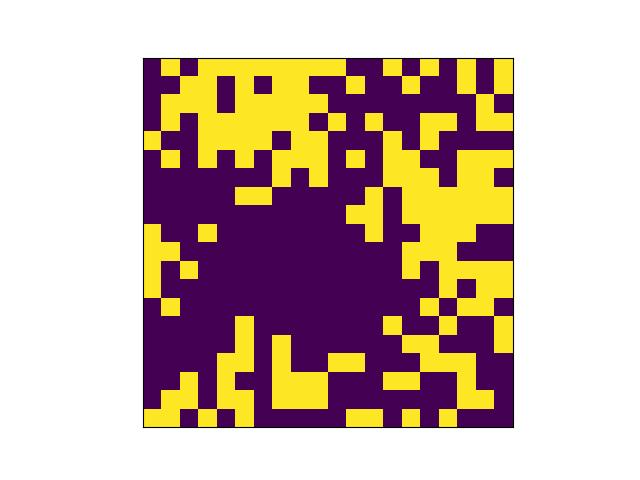}
		\caption{}
	\end{subfigure}
	\hfill
	\begin{subfigure}[b]{0.5\textwidth}
		\includegraphics[width=\textwidth]{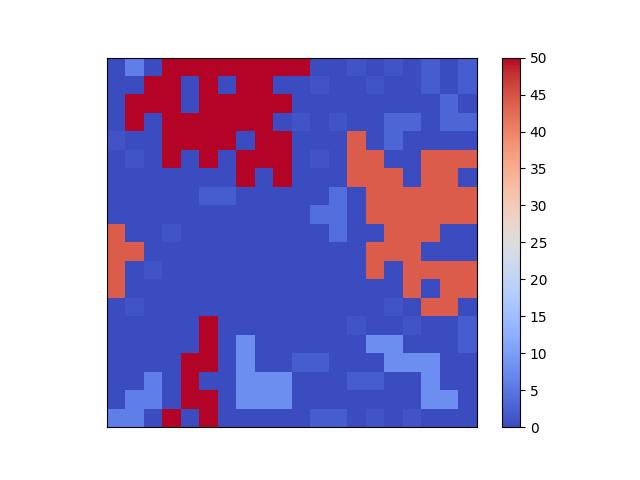}
		\caption{}
	\end{subfigure}
\caption{(a): Spin configuration of the neuronal network at the critical temperature $T = 18$, where each spin represents an active (+1, yellow) or inactive (-1, purple) neuron. (b): Corresponding clusters of active neurons extracted from the configuration in (a), under periodic boundary conditions. The colorbar in (b) indicates the size of each cluster, with brighter colors representing larger clusters.}

	\label{fig:Clusters_area_18}
\end{figure}

A neuron cluster is defined as a group of connected active neurons that are neighbors. Figure b-\ref{fig:Clusters_area_18} shows examples of clusters of different sizes at $T=18$ .

\begin{figure}[htb]
	\centering
	\begin{subfigure}[b]{0.49\textwidth}
		\includegraphics[width=\textwidth]{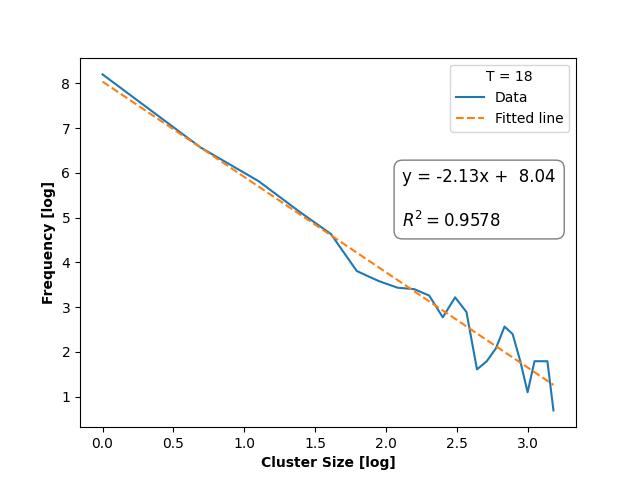}
		\caption{}
	\end{subfigure}
	\hfill
	\begin{subfigure}[b]{0.49\textwidth}
		\includegraphics[width=\textwidth]{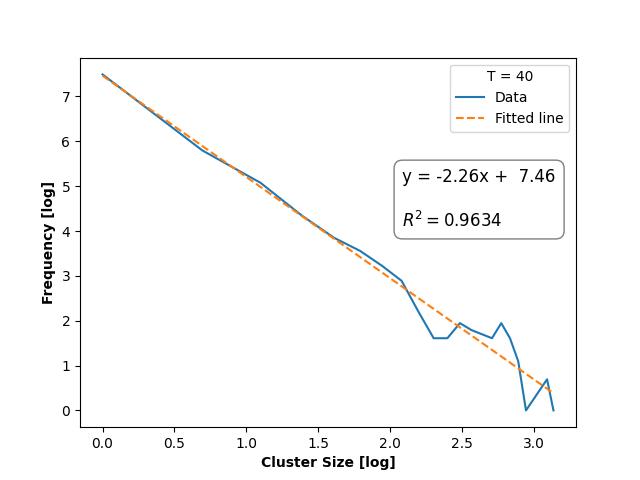}
		\caption{}
	\end{subfigure}
\caption{Cluster size distributions at different temperature regimes. 
	(a) At the critical temperature ($T = 18$), the cluster size distribution follows a power-law with slope $-2.13$, indicating scale-free behavior typical of critical dynamics. 
	(b) At the supercritical temperature ($T = 40$), the distribution also follows a power-law with slope $-2.26$, reflecting oscillatory activation patterns that support the formation of clusters of varying sizes over time.}

	\label{fig:Cluster_Size_Distribution}
\end{figure}

\begin{itemize}
	\item  \textbf{Supercritical regime} ( $T=40$ ): At supercritical temperatures, network  activity shows oscillating on/off behavior (Figure  \ref{fig:raster_mag_plot}). According to Figure (b) \ref{fig:Cluster_Size_Distribution} , the distribution of cluster sizes at $T=40$ is  fitted by a power-law distribution with a slope of -2.26 ($R^2$=0.96). The oscillatory on/off behavior causes clusters of every size to be found in the distribution over time .
	
	\item  \textbf{Critical regime}( $T=18$ ):
	 At the critical temperature ($T=18$), the system is in a phase transition state, and cluster sizes are varied . As seen in Figure b-\ref{fig:raster_mag_plot}, small clusters form quickly and disappear again . Figure a-\ref{fig:Clusters_area_18} shows the on/off neurons at $T=18$ , and Figure b-\ref{fig:Clusters_area_18} shows clusters of different sizes . At the critical temperature, the distribution of cluster sizes fits a power-law with high accuracy, indicating scale-free behavior where clusters of any size can be found .
	 
	 \item  \textbf{Subcritical regime}( $T=10$ ):  At subcritical temperatures, such as $T=10$, according to Figure c-\ref{fig:raster_mag_plot}, most neurons are off, and only a small number activate randomly . Therefore, the concept of a well-defined cluster is practically absent .
\end{itemize}

%%%%%%%%%%%%%%%%%%%%%%

%% file: chapters/conclusions.tex
\section{Conclusion}

We introduced a modified Ising model where each spin acts as a neuron with enforced ON and OFF refractory periods. Simulated on a 2-D lattice with 28-neighbor connectivity using the Metropolis rule, the system shows a continuous phase transition near a critical temperature $T_c \approx 18$, resembling behaviors seen in cortical circuits.

\textbf{Macroscopic behavior:} Mean neuronal activity transitions from asynchronized to synchronized activity as temperature rises, with dynamic capacity and neuron sensitivity peaking at $T_c$, consistent with second-order phase transitions.

\textbf{Temporal and spatial patterns:} Autocorrelation decays slowly at $T_c$, indicating long-term memory. Spatial correlations extend farther at $T_c$, suggesting emergent scale-free structure.

\textbf{Avalanche dynamics:} Neuronal clusters follow power-law distributions  near $T_c$, reflecting critical behavior.

These findings show that minimal biological constraints in an Ising-like model can reproduce critical neural dynamics. Near criticality, systems gain broader stimulus range, flexible dynamics, and efficient communication.

\textbf{Limitations and future work:}
Our model employs a homogeneous lattice, purely excitatory couplings, and fixed on/off times. Extending it to diverse topologies, mixed excitatory–inhibitory interactions, plastic couplings, and biologically measured refractory-period distributions will allow stricter confrontation with \textit{in-vivo} data and with analytical predictions for complex networks. Such refinements should clarify whether real cortical tissue fine-tunes itself to the same balance between stability and sensitivity that we observe here, and could illuminate how disorders or neuromodulation shift the brain away from—or draw it back toward—criticality.